\documentclass[numberedappendix]{emulateapj}
\usepackage{amsmath}
\slugcomment{ApJ, in press; \today}

\newcommand{\G}{\Gamma}

\newcommand{\g}{\gamma}

\newcommand{\psim}{\lower.5ex\hbox{$\; \buildrel \propto \over\sim \;$}}
\newcommand{\lbar}{\lower.0ex\hbox{$\; \buildrel
{\lower0.0ex \hbox{-}} \over\lambda  \;$}}

\newcommand{\tgg}{\tau_{\gamma\gamma}}

\newcommand{\cm}{\mathrm{cm}}

\newcommand{\erg}{\mathrm{erg}}

\newcommand{\MeV}{\mathrm{MeV}}
\newcommand{\GeV}{\mathrm{GeV}}

\newcommand{\s}{\mathrm{s}}

\newcommand{\Gauss}{\mathrm{G}}

\newcommand{\fermi}{{\em Fermi}}

\shorttitle{IGMF Constraints from Gamma Ray Observations}
\shortauthors{Finke et al.}

\begin{document}
\title{Constraints on the Intergalactic Magnetic Field with Gamma-Ray
Observations of Blazars}

\author{Justin D.\ Finke\altaffilmark{1,2}, Luis C.\ Reyes\altaffilmark{3,4}, Markos
Georganopoulos\altaffilmark{5,6}, Kaeleigh Reynolds\altaffilmark{3}, Marco Ajello\altaffilmark{7}, Stephen J.~Fegan\altaffilmark{8}, 
Kevin McCann\altaffilmark{5} }

% maybe include Soebur and Chuck, maybe even Chul.

\altaffiltext{1}{U.S.\ Naval Research Laboratory, Code 7653, 4555 Overlook Ave.\ SW,
        Washington, DC,
        20375-5352, USA}
\altaffiltext{2}{email:  justin.finke@nrl.navy.mil}
\altaffiltext{3}{Department of Physics, California Polytechnic State University, San Luis Obispo, CA 93401, USA}
\altaffiltext{4}{email: lreyes04@calpoly.edu}
\altaffiltext{5}{Department of Physics and Center for Space Sciences and Technology, University of Maryland Baltimore County, Baltimore, MD 21250, USA}
\altaffiltext{6}{email:  georgano@umbc.edu}
\altaffiltext{7}{Department of Physics and Astronomy, Clemson University, Kinard Lab of Physics, Clemson, SC 29634-0978, USA}
\altaffiltext{8}{Laboratoire Leprince-Ringuet, \'Ecole polytechnique, CNRS/IN2P3, Palaiseau, France}

\begin{abstract}

Distant BL Lacertae objects emit $\g$ rays which interact with the
extragalactic background light (EBL), creating electron-positron
pairs, and reducing the flux measured by ground-based imaging
atmospheric Cherenkov telescopes (IACTs) at very-high energies (VHE).
These pairs can Compton-scatter the cosmic microwave background,
creating a $\g$-ray signature at slightly lower energies observable by
the \fermi\ Large Area Telescope (LAT).  This signal is strongly
dependent on the intergalactic magnetic field (IGMF) strength ($B$)
and its coherence length ($L_B$).  We use IACT spectra taken from the
literature for 5 VHE-detected BL Lac objects, and combine it with LAT
spectra for these sources to constrain these IGMF parameters.  Low $B$
values can be ruled out by the constraint that the cascade flux cannot
exceed that observed by the LAT.  High values of $B$ can be ruled out
from the constraint that the EBL-deabsorbed IACT spectrum cannot be
greater than the LAT spectrum extrapolated into the VHE band, unless
the cascade spectrum contributes a sizable fraction of the LAT flux.
We rule out low $B$ values ($B\la10^{-19}$ G for $L_B\ge1$\ Mpc) at
$>5\sigma$ in all trials with different EBL models and data selection,
except when using $>1\ \GeV$ spectra and the lowest EBL
models.  We were not able to constrain high values of
$B$.

\end{abstract}


\section{Introduction}
\label{intro}

The extragalactic background light (EBL) from the infrared (IR),
through the optical and into the ultraviolet (UV) is dominated by
emission from all the stars in the observable universe, either
directly or through dust absorption and reradiation.  It contains
information about the cosmological expansion, star formation history,
dust extinction and radiation in the universe, and so can provide
constraints on a number of cosmologically interesting parameters.
However, its direct detection is hampered by the bright foreground
emission from the Earth's atmosphere, the solar system, and the
Galaxy.  This can be avoided by making measurements from
spacecraft outside the Earth's atmosphere
\citep[e.g.,][]{hauser98,dwek98,bernstein02,mattila03,bernstein07} or
solar system
\citep{toller83,leinert98,edelstein00,murthy01,matsuoka11}, or through
galaxy counts, which in general give lower limits
\citep[e.g.,][]{madau00,marsden09}.  See for example \citet{hauser01} for a review
of EBL measurements, constraints, and models.

Shortly after the discovery of the cosmic microwave background (CMB)
radiation \citep{penzias65}, it was realized that this and other
radiation fields would interact with extragalactic $\g$ rays,
producing electron-positron pairs\footnote{Hereafter, we refer to both
electrons and positrons as simply electrons.}, effectively absorbing
the $\g$ rays \citep{nikishov62,gould67_EBL,fazio70}.  In the
1990s, extragalactic high-energy $\g$-ray astronomy took major leaps
forward, with the launch of the {\em Compton Gamma-Ray Observatory}
and the first detections of extragalactic sources at high energies
(MeV--GeV) by EGRET \citep{hartman92}, and at very-high energies (VHE;
$\ga 0.1$\ TeV) by ground-based imaging atmospheric Cherenkov
telescopes \citep[IACTs; ][]{punch92,mohanty93}.  It was almost
immediately realized that $\g$-ray observations of extragalactic
blazars could be used to constrain the EBL
\citep{stecker92,stecker93,dwek94,biller95,madau96}. The
detection of Mrk 501 out to 20 TeV \citep{aharonian99_mrk501} was not
consistent with the EBL models of the time
\citep{aharonian99_mrk501,protheroe00}.  The detection of the hard
spectrum from the BL Lac 1ES 1101$-$232 with H.E.S.S. seems to put
strong constraints on the EBL, ruling out models that predict high
opacity if one assumes the intrinsic VHE spectrum cannot be harder
than $\G=1.5$, where the differential photon flux $\Phi(E) = dN/dE
\propto E^{-\G}$ \citep{aharonian06}.  However, the $\G=1.5$
constraint has been questioned, and a number of theoretical
possibilities have been raised that could account for harder VHE
spectra \citep*{stecker07_accel,boett08,aharonian08}.  Nonetheless,
the basic idea, that the deabsorbed VHE photon index cannot be harder
than a certain value, has been used by many authors to constrain the
EBL
\citep[e.g.,][]{schroedter05_EBL,aharonian07_0229,mazin07,albert08_3c279,finke09_EBLconstrain}.
EBL constraints from $\gamma$-ray observations have been used to
constrain the contribution of Population II and III stars to the EBL
star formation \citep{raue09,raue12,gilmore12_pop3}, including the
contribution of dark matter-powered stars \citep{maurer12}.

The new era of $\g$-ray astronomy, which has begun with the launch of
the {\em Fermi Gamma-Ray Space Telescope} has brought additional
constraints on the EBL.  The {\em Fermi} Large Area Telescope (LAT) is
sensitive to $\g$ rays from 20 MeV to greater than 300 GeV
\citep{atwood09} and surveys the entire sky every $\sim$ 3 hours,
collecting an unprecedented number of $\g$-ray photons from the entire
sky.  At the energies observed by the LAT, the extragalactic $\g$-ray
sky below $\sim$ 10 GeV is expected to be entirely transparent to $\g$
rays, at least back to the era of recombination ($z\sim 1000$)
\citep[e.g.,][]{oh01,finke10_EBLmodel}, while photons observed in the
10 GeV -- 300 GeV range should be attenuated by UV/optical photons if
they originate from sources at $z\ga0.5$.  This suggests a possible
way to constrain the EBL \citep*{chen04}.  The LAT spectrum below 10 GeV can be
extrapolated to higher energies, and should be an upper limit on the
intrinsic spectrum in the range where the EBL attenuates the $\g$
rays.  This seems to be a reasonable assumption, since no spectrum
that contradicts this has been observed, and it is difficult to
imagine theoretical ways to produce one, although see e.g.
\citet{aharonian02_upturn,aharonian08,lefa11_mrk501}.  \citet{abdo10_EBL} have used
this technique to put upper limits on the optical/UV EBL absorption
optical depth ($\tgg$) for sources $z\ga 0.5$, and rule out with high
significance some models which predict high $\tgg$ with data from the
first 11 months of LAT operation from 5 blazars and 2 gamma-ray bursts
(GRBs).  More recently, \citet{ackermann12} used a similar technique
in a composite fit to 150 BL Lacs with 46 months of LAT data to
constrain the high-$z$ EBL even further, and found agreement with most
recent models
\citep{franceschini08,gilmore09,finke10_EBLmodel,kneiske10,dominguez11,gilmore12_model}.
\citet{abramowski13_hessEBL} performed a combined analysis of
HESS blazar spectra and found them to be consistent with EBL
absorption using the model of \citet{franceschini08}.
\citet{georganopoulos08} suggested the Compton-scattering of EBL
photons by high-energy electrons in the radio lobes of Fornax A could
be detected by the {\em Fermi}-LAT, leading to another possible way to
constrain EBL intensity.

Below 100 GeV, the universe is expected to be transparent out to
$z\sim 0.1$, although VHE photons from this redshift range should be
attenuated by interactions with IR EBL photons.
\citet*{georgan10_EBL} suggested a very similar technique to
\citet{abdo10_EBL}, applied to the VHE range.  For those sources
detected by both LAT and an IACT, one can extrapolate the LAT spectrum
(which should be unattenuated for these sources at low $z$) into the
VHE range, and use it as an upper limit on the intrinsic VHE flux.  As
with the LAT-only case, a comparison of this upper limit with the
observed VHE flux allows one to compute an upper limit on $\tgg$.
\citet{georgan10_EBL} used this technique to show that models which
predicted high $\tgg$ in the VHE range were strongly disfavored.
Another possible way to estimate the intrinsic spectrum of a source,
and thus estimate $\tgg$, comes from modeling the full radio to GeV
$\g$-ray spectral energy distribution (SED) of a $\g$-ray blazar with
a standard synchrotron/synchrotron self-Compton (SSC) model.  The SSC
spectrum can be extrapolated to the VHE regime, and compared with
observations to estimate $\tgg$ \citep{mank11}.  \citet{dominguez13}
have applied this technique to a sample of $\sim 15$ LAT and IACT-detected
blazars to constrain the cosmic $\g$-ray horizon, i.e., the energy
where $\tgg=1$ for a certain redshift.  \citet{dwek13} present a
comprehensive review of recent attempts to constrain the EBL with
$\g$-ray observations.

These constraints come with caveats, however.  A major one is that the
GeV and VHE $\g$ rays must come from the same region in the jet.  But
the VHE $\g$ rays could have a different origin: namely, they could be
created in a different region of the jet \citep{boett08}, or from
ultra-high energy cosmic rays (UHECRs) which originate from the
blazar, and interact with the CMB and IR-UV EBL to produce VHE $\g$
rays \citep{essey10_1,essey10_2,essey11_cr,razzaque12}.  Another
possibility is that the electrons that are produced by the
$\g$-ray--EBL photon interactions Compton-scatter CMB photons,
producing GeV $\g$-ray emission which could itself be absorbed by
interactions with the EBL, producing a cascade
\citep{aharonian94,plaga95,dai02}.  If the intergalactic magnetic
field (IGMF) strength is low, the pairs will not be significantly
deflected from our line of sight, and this could produce an observable
feature in the LAT bandpass \citep[e.g.,][]{neronov09}.  This cascade
component was taken into account in the EBL constraints of
\citet{meyer12_EBL}, and could be an important component that should
be included in the broadband spectral modeling of BL Lac objects
\citep[e.g.,][]{tavecchio11_igmfmodel,tanaka14}.  In recent years
several authors have used the non-detection of these cascades to put
lower limits on the IGMF strength
\citep[e.g.,][]{neronov10,tavecchio10_igmf,dolag11,essey11_igmf}.  In
general these efforts depend on the fact that emission from VHE
blazars is relatively constant over long periods of time ($\ga 10^6$\
years).  Some VHE blazars have been seen to have non-variable emission
on timescales of years \citep[e.g.,][]{aharonian06,aharonian07_0229},
but other blazars are highly variable at these energies
\citep[e.g.,][]{aharonian07_2155}.  Indeed observing and studying the
time-dependent EBL-induced pair cascades from GRBs and blazar flares
has been suggested as a way to probe the IGMF parameters
\citep{razzaque04,ichiki08,murase08}.  The possibility of variable VHE
emission has led to caveats in interpreting the IGMF constraints from
apparently non-variable TeV blazars \citep{dermer11,taylor11}.  Other
uncertainties such as the EBL intensity and VHE spectra errors, blazar
jet geometry and Doppler factor, can further decrease the lower limits
on the IGMF \citep{arlen14}.  For higher values of the
intergalactic magnetic field strength, the cascade emission is not
expected to be a point source, so would result in $\g$-ray ``halos''
in the $\sim 0.1$---$100\ \GeV$ energy range around VHE-detected blazars
\citep{neronov09}.  The detection of these halos was reported by
\citet{ando10_halo}, but it was contested by \citet{neronov11} and
\citet{ackermann13_psf}.  The cascades might also be suppressed by
plasma beam instabilities
\citep{broderick12,schlickeiser12_pair,schlickeiser12_plasma,menzler15},
although this is controversial
\citep{venters13,minati13,sironi14,chang14}.

In this paper we report on our efforts to constrain the IGMF strength
($B_{IG}$) and coherence length ($L_{B}$) using $\g$-ray observations
of blazars from both LAT and IACTs.  This technique can be seen as an
extension of previous work by \citet{georgan10_EBL}, where the focus
was on constraining the EBL.  We make use of data from the first 70
months of LAT operation, the analysis of which is described in Section
\ref{lat_analysis}, and IACT spectra from the literature.  We describe
our technique in section \ref{method} and report on our constraints on
IGMF parameters in Section \ref{results}.  Finally we discuss the
interpretations and implications of our results in Section
\ref{discussion}.

\section{LAT Analysis}
\label{lat_analysis}

To determine the LAT spectra of our sources, listed in Table
\ref{lat_results}\footnote{We describe our source selection in
Section \ref{sourceselection}.}, we considered all LAT data collected
since the start of the science mission in 2008 August 4 (MJD 54682)
until 2014 June 30 (MJD 56838), i.e.\ for 70 months of operation. This
constitutes a significant increase in statistics with respect to
previous efforts \citep[e.g.,][]{neronov10,tavecchio10_igmf}. The data
were analyzed using an official release of the {\em Fermi}
ScienceTools (v9r34p1), \verb P7REP_SOURCE_V15 \ instrument response
functions and considering photons satisfying the \verb SOURCE \ event
selection. We exclude photons detected at instrument zenith angles
greater than 100\arcdeg\ to avoid contamination from the Earth's limb,
and data taken with the Earth within the LAT's field of view by
requiring a rocking angle less than 52\arcdeg.  

For completeness, we created and used LAT spectra both $>0.1$\ GeV and
$>1.0$\ GeV.  The source 1ES 0229$+$200 is significantly contaminated
at low energies by albedo gamma-ray emission from the Moon
\citep{johannesson13}, so for this object we excluded data while the
source was near the Sun or Moon (angular distance less than
$15^\circ$), reducing the exposure by about 10\%.  For all of our
sources, we excluded time intervals around bright gamma-ray bursts and
solar flares, as recommended by \citet{acero15_3fgl}.

The spectral analysis of each source is based on the maximum
likelihood technique using the standard likelihood analysis software.
In addition to Galactic and isotropic diffuse background
components\footnote{The LAT background models are available on the web
at \\
\url{http://fermi.gsfc.nasa.gov/ssc/data/access/lat/BackgroundModels.html}.},
we considered all point-like sources within 15\arcdeg\ from the source
position that we found in the second {\em Fermi} catalog
\citep[2FGL;][]{nolan12_2fgl} and a preliminary version of the third
{\em Fermi} catalog \citep[3FGL,][]{acero15_3fgl}.  Sources within
$5^\circ$ of the source had their normalizations and spectral
parameters (i.e., spectral index and log-parabola width parameter if
appropriate) free to vary.  Sources between $5^\circ$ and $10^\circ$
had their normalizations free to vary but their spectral parameters
fixed to the catalog values.  Sources farther than $10^\circ$ had
their normalizations fixed to their catalog as well.  Any residual
hotspots in the Test Statistic \citep[$TS$;][]{mattox96} maps with $TS
> 10$ and less than $10^\circ$ from the source were modeled as point
sources at the location of the maximum TS value.  A likelihood ratio
test was used to find the best spectral model (power-law, broken power
law, or log parabola) that fit the data. In all cases, there was
insufficient evidence to reject the simple power-law model.  The
power-law spectral parameters resulting from our fits can be found in
Table \ref{lat_results}, including the $TS$, total photon flux
($F_{LAT}$), spectral index ($\G$), and the off-diagonal term in the
covariance matrix \citep[${\rm cov}(F_{LAT}, \G) =
\sigma_{F}\sigma_{\G}\rho$; e.g.,][]{abdo09_tev}.  Using the
off-diagonal terms in the covariance matrix insures that we take into
account the correlation between the errors in the parameters $F_{LAT}$
and $\G$ in our error analysis, described in Section \ref{ruleout}.

In addition to the standard maximum likelihood analysis performed to
find the best fit to the data, the Log-of-the-likelihood ($LL$)
profile as a function of the source's photon flux ($F_{LAT}$) was
calculated in order to fully characterize the uncertainty on this
parameter. While stepping through the source's flux, the other
parameters, such as the source's spectral index, the free parameters
from nearby sources, and the normalizations of the diffuse models,
were left free to vary.  Thus, instead of assuming a perfect normal
(Gaussian) distribution for the error of $F_{LAT}$ and using 1, 2, and
3 standard deviations to calculate the 68\%, 95\% and 99\% confidence
intervals, we used the $LL$ profile to calculate the actual confidence
intervals assuming that $-2\Delta(LL)$ is distributed as the
chi-square probability distribution with one degree of freedom. We
found that this approach is necessary in order to correctly determine
the flux probability distribution of weak LAT sources such as
1ES~0229+200, 1ES~0347$-$121 and 1ES 1101$-$232.  See Section
\ref{pdf_section} for the details of the probability distributions
used in our Monte Carlo (MC) analysis, which is described in Section
\ref{ruleout}.  We also used this technique for the spectral indices
and found that their errors are well-represented by normal
distributions.

\clearpage
%\begin{landscape}
%\LongTables
\begin{turnpage}
\begin{deluxetable}{llccccccccc}
\tabletypesize{\scriptsize}
%\rotate
\tablecaption{Sources and results of LAT analysis.}
\tablewidth{0pt}
\tablehead{
\colhead{Source name} & 
\colhead{3FGL name} & 
\colhead{$z$\tablenotemark{a}} & 
\colhead{VHE instrument} &
\colhead{VHE reference} &
\colhead{LAT energy range} &
\colhead{$TS$} &
\colhead{$F_{LAT}$\tablenotemark{b}} & 
\colhead{$\Gamma$\tablenotemark{c}} &
\colhead{$\rho\sigma_F \sigma_\G$\tablenotemark{d}}  &
\colhead{$V$\tablenotemark{e}} 
}
\startdata
\object{1ES~0229+200} & J0232.8+2016 & 0.139 & HESS & \citet{aharonian07_0229} & 0.1--300 GeV & 58.0 & $1.76\pm0.09$ & $1.71\pm0.16$ & 0.014 & 49 \\
\object{1ES~0229+200} & & & & & 1.0--300 GeV & 52.4 & $0.30\pm0.08$ & $1.62\pm0.17$ & $7.9\times10^{-3}$ & \\
\object{1ES~0347$-$121} & J0349.2$-$1158 & 0.185 & HESS & \citet{aharonian07_0347} & 0.1--300 GeV & 48.8 & $1.18\pm0.49$ & $1.66\pm0.14$ & $5.7\times10^{-3}$ & 44 \\
\object{1ES~0347$-$121} &  &  &  &  & 1.0--300 GeV & 50.2 & $0.30\pm0.07$ & $1.77\pm0.17$ & $3.9\times10^{-3}$ & \\
\object{1ES~0414+009} & J0416.8+0104 & 0.287 & HESS & \citet{abram11_0414} & 0.1--300 GeV & 127 & $3.10\pm0.99$ & $1.74\pm0.10$ & 0.092 & 56 \\
\object{1ES~0414+009} &  &  &  &  & 1.0--300 GeV & 127 & $0.56\pm0.09$ & $1.73\pm0.12$ & $4.7\times10^{-3}$ & \\
\object{1ES~1101$-$232} & J1103.5$-$2329 & 0.186 & HESS & \citet{aharonian07_1101} & 0.1--300 GeV & 75.6 & $1.57\pm0.67$ & $1.63\pm0.14$ & 0.084 & 37 \\
\object{1ES~1101$-$232} &  &  &  &  & 1.0--300 GeV & 70.3 & $0.34\pm0.08$ & $1.59\pm0.15$ & $5.9\times10^{-3}$ & \\
\object{1ES~1218+304} & J1221.3+3010 & 0.182 & VERITAS & \citet{acciari09_1218} & 0.1--300 GeV & 1690 & $15.6\pm1.5$ & $1.68\pm0.03$ & 0.040 & 92 \\
\object{1ES~1218+304} &  &  &  &  & 1.0--300 GeV & 1603 & $32.3\pm1.8$ & $1.69\pm0.04$ & 0.024 & \\
\enddata
\tablenotetext{a}{redshift.}
\tablenotetext{b}{integrated LAT flux in units $10^{-9}$ ph cm$^{-2}$ s$^{-1}$ 
from power-law fit.}
\tablenotetext{c}{Spectral index from LAT power-law fit.}
\tablenotetext{d}{Off-diagonal term of the covariance matrix from the LAT power-law fit in 
units $10^{-9}$ ph cm$^{-2}$ s$^{-1}$ .}
\tablenotetext{e}{Variability Index from the 3FGL.}
\label{lat_results} 
\end{deluxetable} 
%\end{landscape}
\end{turnpage}
\clearpage

\section{Method for Constraining Models}
\label{method}

\subsection{Assumptions}
\label{assumption_section}

Our technique for constraining the IGMF is based on
\citet{georgan10_EBL}, with extensions including a sophisticated 
MC technique to more accurately determine the
significance of the constraints, and the addition of a pair cascade
component.  We make the following assumptions, the first three of
which are identical to those of \citet{georgan10_EBL}:
\begin{enumerate}
\item We assume the MeV-TeV flux from BL Lacs in our sample are
produced cospatially from the sources themselves and not UHECR
interactions in intergalactic space \citep{essey10_1}, and that the
deabsorbed VHE data points never exceed the extrapolated LAT
spectra (within errors).  The latter is a fairly common assumption
made for the purposes of constraining the EBL
\citep[e.g.,][]{georgan10_EBL,meyer12_EBL,sanchez13}, and we
consider it likely, since there is as yet no evidence for concave
upwards $\gamma$-ray spectra.  In the Third LAT AGN catalog
\citep{ackermann15_3lac}, 131 sources significantly prefer a
log-parabola to a power-law spectrum, yet not one has a concave
upwards $\g$-ray spectrum.  See \citet{costamante13} for a different
opinion of this assumption.
\item We assume the objects are not variable at $\g$-ray 
energies within the statistical uncertainties of the measurements.
Indeed, we have selected sources for our sample for which little or no
$\g$-ray variability has been reported, in either the LAT or IACTs.
\item We assume the $\g$ rays will not convert to axion-like particles
or avoid absorption via some other exotic mechanism
\citep[e.g.,][]{deang07,sanchez09,reesman14,meyer14} and that their
absorption is correctly described by the EBL model used.  We use three
EBL models, the ``Model C'' model of \citet{finke10_EBLmodel}, the
lower limit model of \citet{kneiske10}, and the 
model of \citet{franceschini08}.
\item We assume pairs created by $\g$-ray-EBL photon interactions will
Compton-scatter the CMB, and will lose energy primarily through
scattering and not through intergalactic plasma beam instabilities
\citep{broderick12,schlickeiser12_pair,schlickeiser12_plasma}.  Our
technique for including the cascade component is similar to the one
used by \citet{meyer12_EBL}.
\item We assume the IGMF does not change significantly with
redshift.  The IGMF strength is expected to be larger at earlier
redshift as $B_{IG}\propto (1+z)^2$ \citep{neronov09}.  For the low
redshifts probed in this paper, $z<0.3$, the variation in $B_{IG}$
with redshift will be negligible in comparison with the spacing in our
grid of tested values.
\item We assume that all of the cascade flux is produced inside the
LAT PSF.  It is possible the cascade could be extended as observed by
the LAT, due to the pair halo effect \citep[e.g.,][]{neronov09}.
\end{enumerate}

We constrain the IGMF in two ways.  If the cascade flux resulting from
the absorption of VHE $\gamma$ rays exceeds the observed LAT flux, low
values of $B_{IG}$ can be ruled out.  This technique was used
previously by \citet{neronov10} and others.  If $B_{IG}$ is high, the
cascade will be minimal.  If the deabsorbed VHE points are above the
extrapolated LAT spectrum, one could interpret this as ruling out a
particular EBL model, as discussed by \citet{georgan10_EBL}.  However,
this technique neglects the contribution of the cascade flux to the
observed LAT emission.  So we interpret this constraint as ruling out
high $B_{IG}$ values: if the deabsorbed VHE spectral points are above
the extrapolated LAT spectrum, it must mean that there is a
contribution to the LAT spectrum from the cascade component, so that
one cannot know how much of the observed LAT spectrum comes
directly from the source.  A minimum necessary cascade implies an
upper limit on $B_{IG}$.

\subsection{Source Selection}
\label{sourceselection}

\begin{figure*}
%\vspace{2.2mm} 
\vspace{1.5cm} 
\epsscale{1.0} 
\plotone{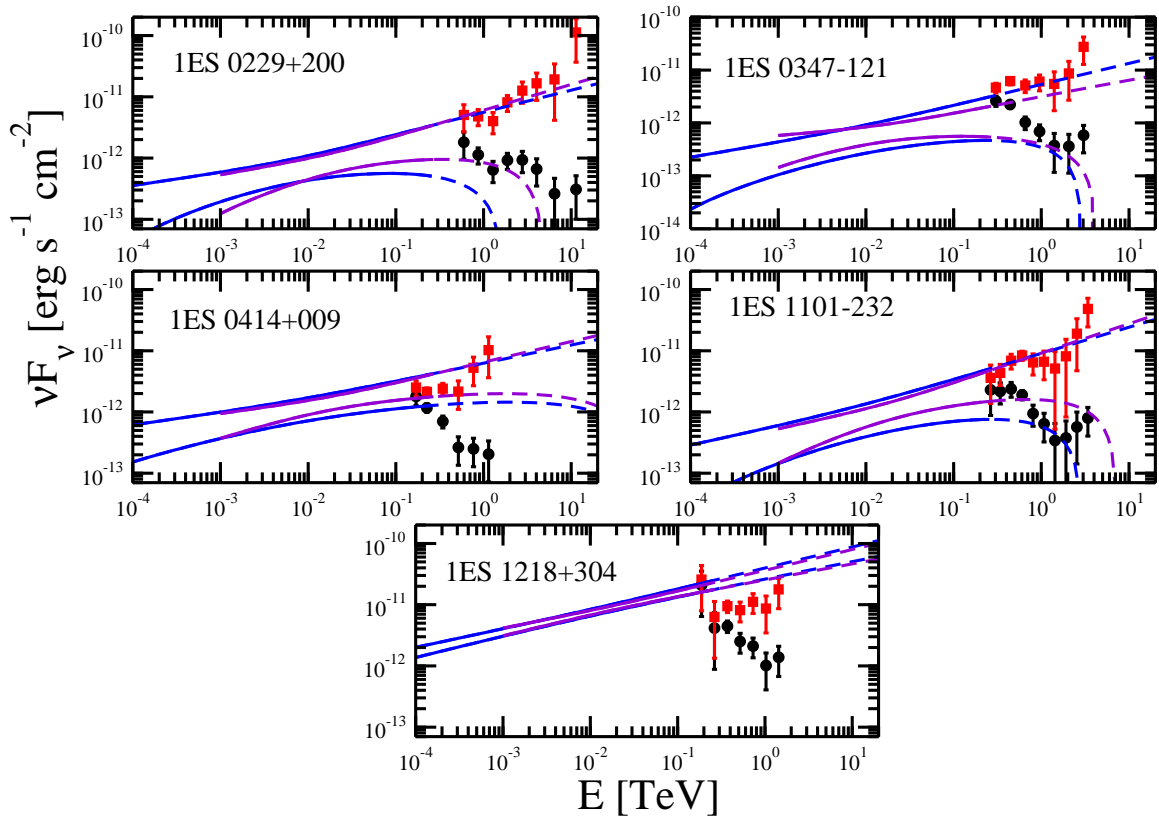}
\caption{ The $\gamma$-ray spectra for the sources.  Blue (violet) curves indicate the 
LAT spectra at $>0.1$ (1.0) GeV, dashed curves indicate the LAT spectra extrapolated 
into the VHE region.  Circles indicate the observed VHE spectrum from ground-based 
atmospheric Cherenkov telescopes, and squares indicate the VHE spectrum deabsorbed with 
the EBL model of \citet{finke10_EBLmodel}.}
\label{spectra}
\vspace{2.2mm}
\end{figure*}

We first selected sources which met the following criteria:
\begin{itemize}
\item We select blazars that are found in the preliminary version of
the 3FGL and in the TeVCat\footnote{The TeVCat can be found on the web
at \url{http://tevcat.uchicago.edu}}.  with a published VHE spectrum
and a measured redshift.
\item We select sources which have variability index $< 100$ in the
3FGL, corresponding to a significance of
$4.8\sigma$\ that the source is variable.  We
prefer sources with little variability, since one of our method's
assumptions is that they have little variability (see Section
\ref{assumption_section}).
\item We restrict ourselves to sources with measured redshift
$z\la0.3$.  This is because the cascade calculation we use, from
\citet{dermer11} and \citet{dermer13_saasfee}, is not valid at
redshifts $z\ga 0.3$.  This excludes some sources which will certainly
be very constraining.  In particular, PKS 1424+240
\citep{acciari10_1424} has its redshift constrained to $z\ga0.6$
\citep{furniss13}, and could be highly constraining.
\item Finally, for the remaining sources, we deabsorbed the VHE
spectra with the EBL model of \citet{finke10_EBLmodel} and plotted
them with the LAT spectra, extrapolated into the VHE regime.  We chose
sources which had deabsorbed VHE spectra above or nearly above their
extrapolated LAT spectra (e.g., Figure \ref{spectra}).  Although
1ES~1218+304 did not meet this criterion, we included it anyway,
because it previously gave strong EBL constraints
\citep{georgan10_EBL}.
\end{itemize}  
Our sources, the details of their LAT spectra, and the sources of
their VHE spectra, are given in Table \ref{lat_results}.

%\clearpage

\subsection{Ruling out Models}
\label{ruleout}

\begin{figure}
\vspace{1.5cm} 
%\vspace{2.2mm} 
\epsscale{1.0} 
\plotone{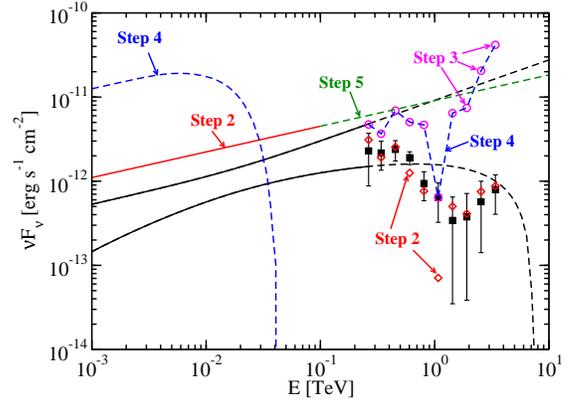}
\caption{ This figure illustrates many of the steps in our method for
ruling out models from Section \ref{ruleout}, using the $\g$-ray
spectrum for 1ES 1101-232.  The LAT spectrum is shown as the bowtie,
along with this spectrum extrapolated to the VHE regime as the dashed
curves.  The observed HESS spectrum is shown as the filled
squares. The randomly drawn HESS points shown as empty diamonds
($F_{VHE}$) and the randomly drawn LAT spectrum is shown as a line,
both of which are labeled ``Step 2''.  The deabsorbed points are shown
as the circles ($F_{VHE,int}$) and labeled ``Step 3''.  The cascaded
component and the interpolated VHE spectrum used to calculate it are
shown as dashed curves labeled ``Step 4''. The LAT spectrum
extrapolated into the VHE regime is shown as the dashed line labeled
``Step 5''.  For the MC iteration shown here, the model is ruled out
by both criterion in Step 5, since $F_{LAT} < F_{cascade}$ and for
several points $F_{LAT,ext} < F_{VHE,int}$.  }
\label{methodfig}
\vspace{2.2mm}
\end{figure}

Our technique for ruling out a model is illustrated in Figure \ref{methodfig} and 
has the following steps:
\begin{enumerate}
\item Select the model we wish to consider.  This includes selecting
an EBL model from the literature, selecting an IGMF strength
($B_{IG}$) and coherence length ($L_B$).  It is also necessary to
select a blazar opening angle.
\item Given the integrated 0.1--300 GeV LAT flux ($F_{LAT}$) and
photon index ($\G$) and their errors for a particular blazar found
with the analysis discussed in Section \ref{lat_analysis}, we draw a
random $F_{LAT}$ and $\G$ from a probability distribution function
(PDF) which represents their errors.  The distribution function used
is discussed in Section \ref{pdf_section}.
\item For each energy bin of the VHE spectrum, with a measured flux
and error, we draw a random flux, $F_{VHE}$, assuming the flux errors
are distributed as a normal distribution (see Section
\ref{pdf_section}).  Each randomly drawn $F_{VHE}(E)$ is deabsorbed
with the EBL model we are testing to give an intrinsic flux,
$F_{VHE,int}(E) = \exp(\tau_{\g\g}(E)) F_{VHE}(E)$.
\item From $F_{VHE,int}(E)$, the contribution of the $e^+ e^-$ pairs
Compton-scattering the CMB, is calculated.  Two of these cascades are
calculated: $F_{cascade,min}$, a cascade with the parameters $t_{\rm
blazar}$ and $E_{max}$ chosen to minimize the cascade; and
$F_{cascade,max}$, with the parameters chosen to maximize the cascade.
These cascades are calculated in the same energy range as the LAT
flux, i.e., 0.1 -- 300 GeV.  Here $t_{\rm blazar}$ is the length of
time the blazar has been emitting $\g$ rays with its current
luminosity \citep{dermer11}, and $E_{max}$ is the maximum energy of
the VHE emission used to calculate the cascade.  Calculating two
cascade components allows us to provide more conservative constraints.
The cascade flux $F_{cascade}$ is calculated using the formula of
\citet{dermer11} and \citet{dermer13_saasfee}.  The minor corrections
to this formula from \citet{meyer12_EBL} should have no effect on the
results, since they affect the lowest flux portion of the cascade
spectrum.  This calculation is compared with MC cascade
calculations from the literature in Appendix \ref{cascade_compare}.
\item The randomly drawn LAT power-law spectrum from step 2 is
extrapolated to the VHE regime.  For this MC iteration, the model is
considered rejected if one of two criteria are met: (i) the minimum
cascade flux from step 4, $F_{cascade,min}$, integrated over the LAT
bandpass exceeds the randomly drawn LAT integrated flux from step 2,
$F_{LAT}$; or (ii) any one of the deabsorbed flux bins from step 2,
$F_{VHE,int}(E)$ exceed the extrapolated LAT flux, $F_{LAT,ext}$, {\em
unless} $0.01 F_{LAT} \le F_{cascade,max}$ in which case the model is
never rejected for this iteration.  If the cascade flux makes up a
significant fraction of the observed LAT flux, we do not believe it
can be extrapolated to the VHE regime and used to constrain models,
since in this case we cannot know the intrinsic spectrum in this
energy range.  The rejection criteria in this step are based on ones
from \citet{georgan10_EBL} and \citet{meyer12_EBL}.
\item Steps 2--5 above are repeated $N_{\rm trials}$ times (we use
$N_{\rm trials}=10^6 $) and the number of times the model is rejected
$N_{reject}$ is counted.  From this, a test statistic is created:
$TS = -2\ln(N_{\rm accept}/N_{\rm trials})$.  Assuming the $TS$ is distributed
as a $\chi^2$ distribution, the p-value for ruling out a model is
calculated.  For models ruled out by the cascade constraint, this
distribution has 2 degrees of freedom; for models ruled out by the VHE
points exceeding the extrapolated LAT spectrum, it has $2N_{VHE}$
degrees of freedom, where $N_{VHE}$ is the number of VHE points.  That
is, the significance for ruling out a model was calculated from
Fisher's method.
\end{enumerate}

In Step 3, each energy bin of the VHE spectra are deabsorbed
assuming the entire bin is at the central energy of the bin.  Although
this could conceivably introduce a systematic error, under the
reasonable assumption that $\tau_{\gamma\gamma}(E)$ is not changing
rapidly within each energy bin, we expect the effect to be
negligible.

In Step 5, the fraction 0.01 was chosen because if the cascade
spectrum contributes less than 1\% to the LAT spectrum, the LAT
spectrum can be safely extrapolated into the VHE range.  This is an
arbitrary choice, but also a very conservative one.  If the cascade
spectrum contributed more than 1\% (say, 10\%) then the LAT spectrum
could still probably be safely extrapolated into the VHE regime.

%\clearpage

\subsection{Probability Distribution Function}
\label{pdf_section}

\begin{figure}
%\vspace{2.2mm} 
\vspace{1.5cm} 
\epsscale{1.0} 
\plotone{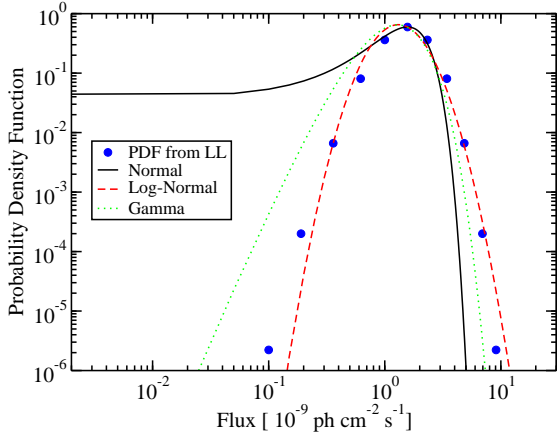}
\caption{ The PDF for the $>100$\ MeV flux for 1ES~1101$-$232 as
determined from the $LL$\ plotted with several
functional forms of the PDF: a Normal distribution, a log-normal
distribution, and a Gamma distribution.  }
\label{pdf1101plot}
\vspace{2.2mm}
\end{figure}

Many of the sources used in our sample are not very bright in the LAT,
and have relatively large errors on their integrated photon flux,
$F_{LAT}$.  Consequently, we found that a bivariate normal probability
distribution did not describe accurately the errors in $F_{LAT}$ and
photon index, $\G$ from the likelihood analysis to the LAT data.  This
PDF faced the problem that there is a significant probability that
$F_{LAT}<0$, which is clearly unphysical.  We plotted the PDF error
points determined from the $LL$ function with several functions
representing the PDF for each source spectrum; an example is seen in
Figure \ref{pdf1101plot}.  Based on these plots, we determined that a
log-normal distribution was a good representation of the PDF for the
$>100\ \MeV$\ spectra, while a Gamma distribution
was a good representation for the $>1\ \GeV$\ 
spectra.  Consequently we used a bivariate normal distribution to
represent the joint PDF for the parameters from the $>100\ \MeV$ LAT
power-law fit, $\Gamma_{LAT}$, and $w=\log_{10}(F_{LAT})$,
\begin{flalign}
\label{pdf_lognormal}
p(F_{LAT}, \G) & = \frac{1}{2\pi\sigma_w\sigma_{\G_{LAT}}\sqrt{1-\rho^2}} 
\nonumber \\  & \times
\exp\Biggr\{ -\frac{ (w-\mu_w)^2}{ 2 (1-\rho^2) \sigma_w^2 }
             -\frac{(\G_{LAT}-\mu_{\G_{LAT}})^2}{2(1-\rho^2)\sigma_{\G_{LAT}}^2}  
\nonumber \\  & 
           + \frac{ \rho(w-\mu_w)(\G_{LAT}-\mu_{\G_{LAT}})}{(1-\rho^2)\sigma_w\sigma_{\G_{LAT}}} 
	     \Biggr\} \ 
\end{flalign}
where $\mu_w$ ($\mu_\G$) is the measured $\log_{10}(F_{LAT})$ ($\G$),
$\sigma_w$ ($\sigma_\G$) is the standard error from the power-law fit
to the LAT data for $\log_{10}(F_{LAT})$ ($\G$), and $\rho$ is the
correlation coefficient between $\log_{10}(F_{LAT})$ and $\G$ (i.e.,
the covariance is $\rho \sigma_w \sigma_\G$).  The error in
$\log_{10}(F_{LAT})$ is calculated from the error in the flux
($\sigma_{F_{LAT}}$) with
$\sigma_w=\sigma_{F_{LAT}}/(F_{LAT}\ln(10))$.  

%\clearpage

For the $>1\ \GeV$\ spectra, where a Gamma distribution best
represented the flux from the power-law fit, we created an ad hoc
bivariate probability distribution based on a Gamma distribution.  It
is given by
\begin{flalign}
\label{pdf_gammanormal}
p(F_{LAT}, \G) & = \frac{ F_{LAT}^{\alpha_F-1} 
e^{-F_{LAT}/\beta_F} }{\G_f(\alpha_F)\beta_F^{\alpha_F} } \ 
\frac{1}{\sigma_{\G} \sqrt{2\pi(1-\rho)} } \ 
\\ \nonumber & \times
\exp\Biggr\{ -\frac{(\G-\mu_\G)^2}{2(1-\rho^2)\sigma_{\G}^2} 
\\ \nonumber & 
            + \frac{ \rho(F_{LAT}-\mu_F)(\G-\mu_\G)}{(1-\rho^2)\sigma_F\sigma_\G} 
\\ \nonumber & 
	    - \frac{ \rho^2(F_{LAT}-\mu_F)^2}{ 2 (1-\rho^2) \sigma_F^2 } \Biggr\} \ ,
\end{flalign}
where $\mu_F$ ($\mu_\G$) is the measured $F_{LAT}$ ($\G$), and
$\sigma_F$ ($\sigma_\G$) is the standard error from the power-law fit
to the LAT data for $F_{LAT}$ ($\G$), $\alpha_F=(\mu_F/\sigma_F)^2$,
$\beta_F=\sigma_F^2/\mu_F$, $\rho$ is the correlation coefficient
between $F_{LAT}$ and $\G$ (i.e., the covariance is $\rho \sigma_F
\sigma_\G$), and
\begin{flalign}
\G_f(x) = \int_0^{\infty}\ dy\ e^{-y}\ y^{x-1}\ 
\end{flalign}
is the standard Gamma function.  This distribution was constructed to
resemble a Gamma distribution for $F_{LAT}$, a normal distribution for
$\G$, preserve the correlated errors of a bivariate normal
distribution, be normalized to unity, and reduce to a bivariate normal
distribution for $\alpha_F \gg 1$.  The latter two properties are
explored in Appendix \ref{pdfappendix}.  This bivariate PDF is
not unlike the gamma-normal distribution explored by
\citet{alzaatreh14}, although their distribution does not preserve the
correlation between the two variables, and so is not useful for our
purposes.

The errors on the flux of each VHE bin is assumed to be described 
by a normal (Gaussian) distribution, given by 
\begin{flalign}
\label{normaldist}
p(F_{VHE}) = \frac{1}{\sigma_{F_{VHE}} \sqrt{2\pi}} 
\exp \left[ -\frac{1}{ 2\sigma_{F_{VHE}} }( F_{VHE}-\mu_{F_{VHE}} )^2 \right]
\end{flalign}
where $F_{VHE}$ is the randomly drawn flux in that energy bin, and $\mu_{F_{VHE}}$
and $\sigma_{F_{VHE}}$ are the reported flux and measurement error, respectively.

\subsection{Combined Constraints}

For a given model (Step 1 in Section \ref{ruleout}), we wish to
combine the constraints from all of our objects (Table
\ref{lat_results}) to provide the strongest constraint possible for a
given model.  Since the results for different objects are entirely
independent, this is done with Fisher's method 
\citep{fisher25,mosteller48}.  We first created a test statistic for
all the sources,
$$
TS=-2\sum_{k=0}^{N_{s}}\ln(P_{\rm accept,k})
$$ from the individual p-values for each source, $P_{\rm accept,k}$,
where $N_s$ is the number of sources.  Fisher's method assures
that the $TS$ is distributed as a $\chi^2$ distribution with $2N_s$
degrees of freedom.  This $\chi^2$ distribution is integrated, giving
the overall p-value of acceptance, $P_{\rm accept,com}$.  We choose to
present the combined results for rejecting a model as the equivalent
number of sigma the model is rejected if the error were distributed as
a normal distribution.  That is, the number of sigma a model is
rejected is $\Sigma = \sqrt{2}\ {\rm erf}^{-1}(P_{\rm accept,com})$.

\section{Results}
\label{results}

\subsection{Results with Conservative Assumptions}
\label{conservative}

Here we show the results for our conservative assumptions.  We choose
a jet opening angle of $\theta_j=0.1$\ rad, roughly consistent with
values from VLBI measurements \citep{jorstad05}, and the EBL model
from \citet[][their ``model C'']{finke10_EBLmodel}.  For calculation
of $F_{cascade,min}$ we use $t_{\rm blazar}=3$\ years and $E_{max}$ equal
to the central energy of the maximum observed bin from the IACTs.  This
$t_{\rm blazar}$ is the typical time between observations for the objects
in our sample, and the typical time for which we know the sources are
not variable.  For calculation of $F_{cascade,max}$ we use
$t_{\rm blazar}=1/H_0$, i.e., we assume the blazar has been emitting VHE
$\g$ rays at the level currently observed for the entire age of the
universe; and $E_{max}=100$\ TeV.  For calculation of
$F_{cascade,max}$ the deabsorbed VHE points are fit with a power-law
and extrapolated to 100 TeV to calculate the cascade component.  The
VHE spectrum is assumed to have a hard cutoff at $E_{max}$.  That is,
this assumes the source does not emit {\em any} $\g$ rays above
$E_{max}$.

\begin{figure}
\vspace{2.2mm} 
\epsscale{1.0} 
\plotone{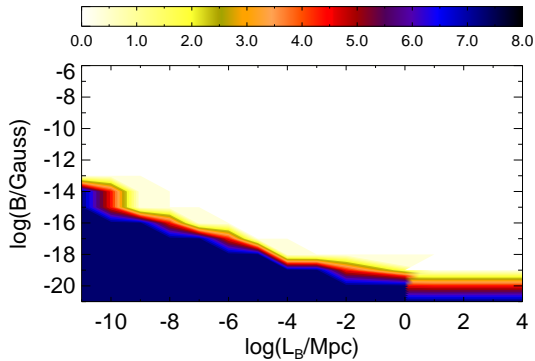}
\caption{ The values of parameter space of $B$ and $L_B$ ruled out for
the combined {\em conservative} results of Section \ref{conservative}
for all of our objects.  The contours represent the significance a
particular region of parameter space is ruled out, in number of sigma,
as indicated by the bar.  These constraints assume the
\citet{finke10_EBLmodel} EBL model and $\theta_j=0.1$ rad.  }
\label{result_finke}
\vspace{2.2mm}
\end{figure}

Our conservative results can be seen in Figure \ref{result_finke}.
One can see that high magnetic field values ($B\ga 10^{-12}$\ G for
$L_B\ga 1$\ Mpc) are not significantly ruled out, while low values
($B\la 10^{-16}$\ G at $10^{-10}$\ Mpc; $B\la 10^{-21}$\ G for $L_B\ga
1$\ Mpc) are ruled out at $\approx7.2\sigma$. For $L_B\ga1$\ Mpc, the
allowed $B$ is essentially independent of $L_B$, since above this
$L_B$ the electrons will lose most of their energy from scattering
within a single coherence length.  For $L_B\la1$\ Mpc, the allowed $B$
goes as $B\propto L_B^{-1/2}$ due to the random change in direction of
$B$, and hence the direction of the electrons' acceleration, as they
cross several coherence lengths.  This overall dependence of the
constraints on $B$ and $L_B$ has been pointed out previously by
\citet{neronov09} and \citet{neronov10}.  There is a strange shape in
the contours at $1-10$\ Mpc due to this transition region, and due to
the coarseness of our grid, which is one order of magnitude in both
$B$ and $L_B$.

Low magnetic field values are inconsistent with the data at
$>5\sigma$.  We consider this quite a significant constraint.  Since
many authors \citep[e.g.,][]{neronov10,dermer11} have ruled out low
$B$ values if the cascade component is above the LAT $2\sigma$ upper
limits, those authors are implicitly ruling out the $B$ values at the
$2\sigma$ level.  The high magnetic field values are not
significantly ruled out.  The most constraining sources in our sample
for low $B$ values turned out to be 1ES~0229+200, 1ES~0347$-$121,
and 1ES~1101$-$232, all of which individually ruled out low $B$ values
at $\ga4.5\sigma$.

Our lower limits on $B$ are lower than what many previous authors have
found in a similar fashion, but assuming $t_{\rm blazar}=1/H_0$
\citep[e.g.][]{neronov10, tavecchio10_igmf,tavecchio11_igmfmodel,
dolag11}.  We compute a constraint with this less conservative
assumption on $t_{\rm blazar}$ below in Section \ref{less_conserv} for
comparison.  Several authors have constrained the IGMF to be $B\ga
10^{-18}$\ G for $L_B=1$\ Mpc by using a shorter $t_{\rm blazar}$ as
we do \citep[e.g.,][]{dermer11,taylor11,vovk12}.  Our lower limits are
generally consistent with these authors, although slightly lower
($B>10^{-19}$\ G).  The minor difference could be due to the fact that
we assume a sharp cutoff at high energies in the intrinsic
spectrum at the maximum VHE energy bin observed from a source, while
other authors extrapolate above this energy in some way, typically
with an exponential form.  This makes our results more conservative.

%\clearpage

\subsection{Robustness}
\label{robustness}

In general, we consider our assumptions, and the results found in
Section \ref{conservative} quite reasonable, and indeed quite
conservative.  However, to be thorough, we have tested the robustness
of these results by varying some of the assumptions, particularly
those that would weaken the constraints, and seeing if this made a
significant difference in our results.

\begin{figure}
\vspace{2.2mm} 
\epsscale{1.0} 
\plotone{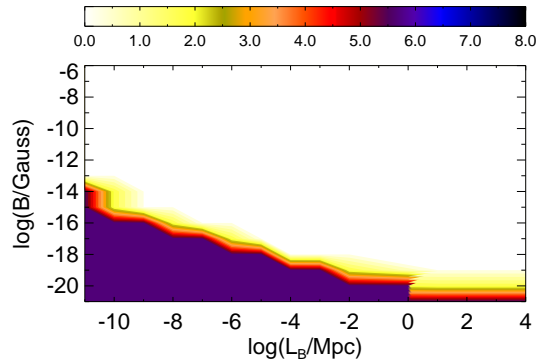}
\caption{ The same as Figure \ref{result_finke}, only with
the EBL model of \citet{kneiske10}. }
\label{result_kneiske}
\vspace{2.2mm}
\end{figure}

The first item we explored is the EBL model.  One would expect
that the parameter space will be ruled out with greater significance if a more
intense and absorbing EBL model is used, while it would be ruled out
with lesser significance if a less intense EBL model is used.  We
performed simulations for a less intense EBL model, namely the model
of \citet{kneiske10}.  This model was designed to be as close as
possible to the observed lower limits on the EBL from galaxy counts;
however, note that for some regions of parameter space, other EBL
models predict less absorption.  The results can be seen in Figure
\ref{result_kneiske}.  The low $B$ values are ruled out at
$5.5\sigma$, while the high $B$ values are still unconstrained.  
We also performed simulations with the model of
\citet{franceschini08}, which has a similar overall normalization as
the \citet{finke10_EBLmodel} model, but its SED has a bit different
shape.  With this model we found that low $B$ values are ruled
out at $6.7\sigma$, and high $B$ values are again
unconstrained.

%\clearpage

\begin{figure}
\vspace{2.2mm} 
\epsscale{1.0} 
\plotone{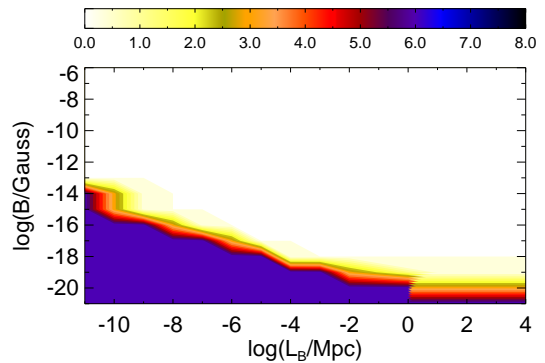}
\caption{ The same as Figure \ref{result_finke}, only without the 
results from the source 1ES~0229+200 and 1ES~1218+304, which have shown 
evidence for $\gamma$-ray variability. }
\label{result_no0229}
\vspace{2.2mm}
\end{figure}

There is some evidence in recent years that the source 1ES~0229+200 is
variable at VHE energies \citep{aliu14_0229}, as is 1ES~1218+304.
We have therefore computed our constraints leaving out these sources,
and the results can be seen in Figure \ref{result_no0229}.  Similar
regions of parameter space are ruled out, but at much less
significance; low values of $B$ are ruled out at $6.0\sigma$.

%\clearpage

We performed simulations with both larger ($\theta_j=0.2$\ rad) and
smaller ($\theta_j=0.05$\ rad) values of the jet opening angle.  A
Larger value of $\theta_j$ led to larger cascades, and increased
significance for ruling out the lower $B_{IG}$ values, but a decreased
significance for ruling out the larger $B_{IG}$ values.  Smaller
values led to smaller cascades, and a decreased significance for
ruling out the lower $B_{IG}$ values, and an increased significance
for ruling out the larger $B_{IG}$ values.  However, we found in all
cases tested that the different jet opening angles made a minuscule
difference, at the level of hundredths of a sigma.

We have so far used the $>0.1$\ GeV LAT spectra for all sources.  If
we use $> 1$\ GeV spectra for all sources, making sure we use
Equation (\ref{pdf_gammanormal}) for the PDF, we find the low values
of $B$ are ruled out at $6.2\sigma$.

Our method is sensitive to the highest energy points from an
IACT spectrum and this could lead to a bias if such points 
are overestimated.  Therefore, we compute 
our constraints throwing away the highest two VHE energy bins.  
In this case, the significance of ruling out the low $B$ values
decreases to $5.9\sigma$.

We note that a log-normal distribution does not
perfectly describe the LAT PDF for the $>100\ \MeV$\ spectra, so we
also performed our simulations with the LAT PDF described by Equation
(\ref{pdf_gammanormal}) rather than Equation (\ref{pdf_lognormal}).
This increased the significance of ruling out the lower $B$ values
slightly, to $7.6\sigma$.

There is some evidence of inaccuracies in the cross-calibration
between the LAT and IACTs, on the order of 10--15\% \citep{meyer10}.
We performed our simulations with the VHE points scaled down by 10\%
and find that this decreases the significance for ruling out low $B$
values to $6.9\sigma$.

\citet{arlen14} use a $\chi^2$ fitting technique to LAT and VHE data
with a primary plus cascade model with IGMF strength $B=0$, and use
the quality of the fits to determine if the data are consistent with
$B=0$.  None of their spectra extend down to 100 MeV; thus the spectra
we use that most resembles theirs is our $> 1$\ GeV spectra.  For
1ES~0229$+$200, for two EBL models, which most resemble the model of
\citet{kneiske10}, they found that the data are not consistent with
$B=0$ at the $>2\sigma$ level.  This is consistent with our results,
since we found for all of our EBL models that $B>0$ at $>3.5\sigma$
for this source with $>1$\ GeV spectra.  \citet{arlen14} also do a fit
with an EBL model with intensity less than 2 nW m$^{-2}$ s$^{-1}$ in
the 10-20 $\mu$m range (their ``EBL model 3'') and find their data are
consistent with $B=0$ at the $2\sigma$ level.  This EBL model is below
all of the EBL models explored here, and indeed it is inconsistent
with 15 $\mu$m galaxy count lower limit from Infrared Space
Observatory observations \citep{metcalfe03}.  For 1ES~0347$-$121, and
1101$-$232, \citet{arlen14} find that they cannot rule out $B=0$ at
$>2\sigma$ with their technique.  With our technique, for the
\citet{kneiske10} EBL model and using $>1$\ GeV LAT spectra, we are
also not able to rule out these sources at $>2\sigma$, so that these
results are also consistent with those of \citet{arlen14}.  Combining
the constraints for all sources with the \citet{kneiske10} EBL model
and the $>1$\ GeV LAT spectra, lower $B$ values (including naturally
$B=0$) are ruled out at only $2.6\sigma$.  This is greater than the
$2\sigma$ criterion of \citet{arlen14}, but a much lower significance
than the $>5\sigma$ constraints from our other assumptions.  We also
found that using the EBL model of \citet{franceschini08}, low $B$
field values were ruled out at $4.8\sigma$ combining the results of
all of our sources.  Ruling out low $B$ field values at $>5\sigma$
significance depends on using $>100$\ MeV spectra and the EBL model
used.

\subsection{Results with Less Conservative Assumptions}
\label{less_conserv}

We have also computed constraints with less conservative assumptions.
We did this by modifying step 4 in our procedure (Section
\ref{method}) as follows: we used $F_{cascade,min}=F_{cascade,max}$,
where both were calculated assuming $t_{\rm blazar}=1/H_0$ and using the
maximum observed VHE photon bin for $E_{max}$.  These less
conservative constraints are computed here primarily for comparison
with other authors, who have used similar assumptions to constrain the
IGMF \citep[e.g.][]{neronov10, tavecchio10_igmf,
tavecchio11_igmfmodel, dolag11}.  

\begin{figure}
\vspace{2.2mm} \epsscale{1.0}
\plotone{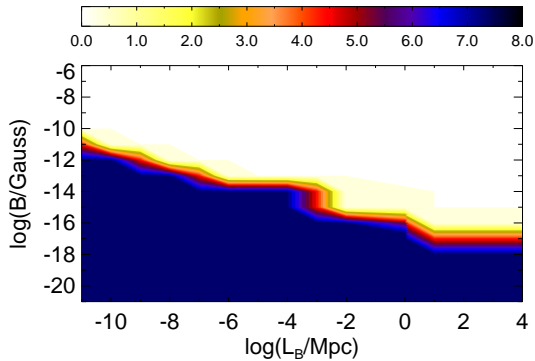}
\caption{ The same as Figure \ref{result_finke}, only with less
conservative assumptions.  Here $F_{cascade,min}=F_{cascade,max}$, and
the cascade was calculated assuming $t_{\rm blazar}=1/H_0$ and $E_{max}$
is the maximum observed VHE photon bin from the source.}
\label{result_notconserv}
\vspace{2.2mm}
\end{figure}

The less conservative constraints can be seen in Figure
\ref{result_notconserv}.  The constraints are similar to the
conservative constraints (Figure \ref{result_finke}), with similar
significance of rejection; however, a much greater region of parameter
space is ruled out.  In particular, $B$ is constrained to be 2-3
orders of magnitude higher than the conservative constraint.  Indeed,
for a given value of $L_B$, $B$ is constrained to within two orders of
magnitude.  For instance, for $L_B=1$\ Mpc, the magnetic field
strength is constrained to $B\ga 10^{-16}$\ G at
$7.1\sigma$.  In general, the lower limit is
roughly consistent although still lower than what previous authors
have found, who generally constrain $B\ga 10^{-15}$ G at $L_B=1$ Mpc
\citep{tavecchio10_igmf,dolag11} although \citet{neronov10} found a
lower limit of $B=3\times10^{-16}$\ G at $L_B\ga1$\ Mpc.  The reasons
for this minor disagreement are probably the same as those discussed
in Section \ref{conservative}.

\citet{essey11_igmf} have computed both upper and lower constraints on
the IGMF assuming $\g$ rays originate from the blazar, or
alternatively, from UHECRs originating from the blazar interacting
with the EBL and CMB.  Only their results for the assumption of $\g$
rays directly from the source are directly comparable to ours, since
we also make this assumption.  They find constraints for various
spectral indices of the intrinsic spectrum.  Their lower limits on $B$
were computed in a similar fashion to other authors
\citep[e.g.,][]{neronov10,tavecchio10_igmf} using the LAT upper limits
and assuming the cascade emission could not exceed this.  They also
calculated upper limits on $B$, because they found that the cascade
flux was necessary to explain the lower energy emission observed by
the IACTs.  For a low EBL model and a high energy cutoff in the
intrinsic spectrum at 20 TeV, they find that essentially no value of
$B$ is allowed for $L_B=1$\ Mpc.  For higher EBL models and higher
energy cutoffs, they did find values of $B$ that were allowed,
generally $B\ga10^{-16}$\ G, and upper limits strongly dependent on
the assumed intrinsic spectral index.  Our constraints are
considerably weaker, and indeed, more conservative.  The main reason
seems to be that we cut off the VHE spectrum at the highest observed
energy bin, rather than assuming it extends to 20-100 TeV.  This means
the cascade flux will be lower, and it will not extend into the VHE
range.  Thus the criterion \citet{essey11_igmf} used to obtain upper
limits on $B$ are not applicable here.

%\clearpage

\section{Discussion}
\label{discussion}

We have used a combination of up-to-date {\em Fermi}-LAT observations
and archival VHE IACT observations of 5 BL Lac objects to constrain
the IGMF parameters $B$ and $\lambda_B$.  These constraints rely on
the assumptions outlined in Section \ref{assumption_section}.  Our
results indicate that magnetic field strength values $B\la 10^{-19}$\
G for $L_B\ge1$\ Mpc are ruled out at $>5\sigma$, with higher $B$
values ruled out for lower values of $L_B$, for example, if
$L_B\approx 10^{-8}$\ Mpc, $B\la10^{-15}$\ G.  These results are
robust with respect to the choice of EBL model, VHE variability of the
sources, jet opening angle, or data selection, i.e., whether $>100$\
MeV or $>1$ GeV LAT spectra are used, and whether or not the highest
two VHE energy bins are used.  The only exceptions are
using the $> 1\ \GeV$ spectra with the EBL models of \citet{kneiske10}
and \citet{franceschini08}.  Using the former EBL model, low $B$
values are ruled out at $2.6\sigma$, and using the latter EBL model
they are ruled out at $4.8\sigma$.  We were not able to constrain high
values of the IGMF, despite our efforts with a new method for doing
so.  We used several novel techniques in our analysis, including a MC
scheme and new bivariate probability distribution to describe the
errors in the LAT power-law fit.  In a preliminary analysis using 42
months of data, we \citep*{finke13} found that large values of $B$
were ruled out at $>5\sigma$ using 1ES~0229+200 and 1ES~1101$-$232.
We were unable to verify this with more data and our updated analysis
(more nearby point sources, exclusion of the Sun and Moon see Section
\ref{lat_analysis}) and updated method for ruling out models (Section
\ref{ruleout}).

These results have several possible interpretations.  The most
straightforward is to take the constraints in Figure
\ref{result_finke} at face value.  In this interpretation, low
magnetic field values are highly unlikely, at the $>5\sigma$ level,
while high magnetic field values are unconstrained.  In all cases, the
VHE spectra are consistent with the extrapolated LAT spectra.  If all
our assumptions are correct, and the constraints can be taken at face
value, our results have implications for the formation of the IGMF and
the era of inflation.  A non-helical IGMF generated from an electroweak
phase transition in the radiation dominated era is essentially ruled
out \citep{neronov09,wagstaff14}.  The BICEP2 collaboration has
claimed the detection of inflation-originating gravitational waves in
the CMB polarization \citep{ade14}.  If the IGMF originates from
inflationary magnetogenesis, there is some tension between the BICEP2
result and constraints for $B\ga10^{-15}$\ G
\citep{fujita12,fujita14,ferreira14}.  However, a joint analysis by
the {\em Planck} and BICEP2/{\em Keck Array} Collaborations
\citep{ade15} found that the original BICEP2 analysis did not properly
take into account a dust contribution to the observed polarization, so
that the claimed detection of gravitational waves was in error.

We do not find any evidence for a contribution from cascade emission
to the LAT flux of our sources. Thus, another possible interpretation
is that plasma beam instabilities eliminate the cascade \citep[][see
assumption \#4 in Section \ref{assumption_section}]{broderick12}.  See
\citet{venters13}; \citet{minati13}; and \citet{sironi14} for
critical assessments of the application of plasma beam instabilities
to VHE blazars.  \citet{chang12} argue that blazars' VHE emission is
absorbed by EBL interactions, and the resulting pairs interacting via
this plasma beam instability contribute to the heating of the
intergalactic medium.  This in turn would prevent the formation of
dwarf galaxies at late times ($z\la2$), possibly reconciling the
sparsity of dwarf galaxies with respect to predictions of $\Lambda$CDM
cosmology \citep{pfrommer12}.

We assume the LAT and VHE $\gamma$ rays are produced co-spatially in
the jet of the BL Lac object (Assumption \#1).  However, it is
possible that UHECRs produced by the object escape into intergalactic
space, where they produce the VHE $\gamma$ rays observed with IACTs
\citep{essey10_1,essey10_2,essey11_cr}.  If this is the case, our
results will not be valid, although it is still possible to constrain
the IGMF \citep{essey11_igmf}.  Since a UHECR origin for VHE $\gamma$
rays would not predict any variability, if the objects used here were
to be observed in the future with an IACT to be at a flux level much
lower than measured previously, this would be strong evidence for the
co-spatially of the LAT and IACT-detected $\gamma$ rays.

If axion-like particles existed with the relevant parameters, the VHE
$\g$ rays could avoid EBL attenuation \citep[assumption
\#3;][]{deang07,sanchez09,horns12,reesman14,meyer14}.  This would allow VHE
photons to avoid much of the $\g$-ray attenuation, meaning our
de-absorbed VHE spectra would not be correct, and the cascade
component would be lower.  This would mean our $B$ constraints are not
valid.  If we had significantly ruled out high magnetic field values,
this could have been interpreted as evidence for this mechanism for
avoiding $\g$-ray attenuation.

The assumption that all of the cascade flux is produced 
inside the PSF (assumption \#6) should not affect our low $B$ 
constraints, since the extended halos should only be observable for 
higher $B$ values \citep{neronov09}.  It is possible that the 
reason we cannot constrain the highest $B$ values is that the cascade 
emission is not present because it is outside the PSF.  

Therefore, one of the most straightforward ways to confirm our results
would be the detection of resolved $\g$-ray halos by LAT or the future
Cherenkov Telescope Array (CTA).  These halos are expected to be
resolved by the LAT if $10^{-16}\la B\la 10^{-14}$\ G for $L_B\ge 1$\
Mpc, and at higher values of $B$ for lower $L_B$ \citep{neronov09}.
We note that both our conservative (Figure \ref{result_finke}) and
non-conservative (Figure \ref{result_notconserv}) constraints allow
for this region of parameter space.  The detection of these halos
would essentially imply that all of our assumptions (Section
\ref{assumption_section}) are correct, except possibly assumption \#1.
It is possible that UHECR interactions with the EBL and CMB could also
produce these $\g$-ray halos.  The IGMF could also be constrained by
observations of anisotropy in the extragalactic $\gamma$-ray
background \citep{venters13}.  \citet{tashiro14} and
\citet{chen15} find a preference for photons of decreasing energy in
the LAT extragalactic $\g$-ray background to be preferentially bent to
the left.  They interpret this as evidence for a helical IGMF with
$B\sim10^{-14}\ \Gauss$ and $L_B\sim10$\ Mpc, which would be
consistent with our results.

\acknowledgements 

%Acknowledge: Discussions with Khota Murase, Warren Essey, Christopher
%Pfrommer, Chul Gwon (hpc), Eileen Meyer (statistics), Kevin McCann (statistics), 
%Tyrel Johnson (LAT Analysis), 
%, HPC at NRL, others.
% add Chuck, Soebur and Marco Ajello here if they aren't included as authors.
% Yasu Tanaka for 0347.

The authors are grateful to the anonymous referee for a prompt,
thorough, and constructive report.  JDF would like to thank Charles Dermer,
Warren Essey, Khota Murase, Christoph Pfrommer, and Soebur Razzaque
for discussions on constraining the EBL and IGMF.  The authors would
also like to thank Pascal Fortin for performing some of the early LAT
analysis on 1ES~0229+200, Tyrel Johnson for discussions on LAT
analysis, Eileen Meyer and Aneta Siemiginowska for discussions on
statistics, Yasuyuki Tanaka for bringing the source 1ES 0347$-$121 to
our attention as a potential source for constraining the IGMF, and
Jeremy Perkins and Dario Gasparrini for thoroughly proofreading the
manuscript.  JDF was supported by the Chief of Naval Research.  This
work was partially supported by a grant of computer time from the
Department of Defense High Performance Computing Modernization Program
at the Naval Research Laboratory.

The \textit{Fermi} LAT Collaboration acknowledges generous ongoing
support from a number of agencies and institutes that have supported
both the development and the operation of the LAT as well as
scientific data analysis.  These include the National Aeronautics and
Space Administration and the Department of Energy in the United
States, the Commissariat \`a l'Energie Atomique and the Centre
National de la Recherche Scientifique / Institut National de Physique
Nucl\'eaire et de Physique des Particules in France, the Agenzia
Spaziale Italiana and the Istituto Nazionale di Fisica Nucleare in
Italy, the Ministry of Education, Culture, Sports, Science and
Technology (MEXT), High Energy Accelerator Research Organization (KEK)
and Japan Aerospace Exploration Agency (JAXA) in Japan, and the
K.~A.~Wallenberg Foundation, the Swedish Research Council and the
Swedish National Space Board in Sweden.
 
Additional support for science analysis during the operations phase is
gratefully acknowledged from the Istituto Nazionale di Astrofisica in
Italy and the Centre National d'\'Etudes Spatiales in France.

\appendix

\section{Comparison of Cascade Calculation with Calculations from the 
Literature}
\label{cascade_compare}

We have made use of the simple analytic cascade calculation described
by \citet{dermer11} and \citet{dermer13_saasfee}.  This calculation
has the advantage of being relatively quick to calculate numerically.
In this appendix, we compare it to more extensive MC cascade
calculations from the literature, those from \citet{taylor11},
\citet{kachelreiss12}, and \citet{arlen14}.  We will use cascades
calculated for 1ES~0229+200, a common source used for these sorts of
constraints.  The primary $\nu F_\nu$ spectrum from the source in all
cases can be parameterized as
\begin{flalign}
F(E) = F_0 \left( \frac{E}{E_0}\right)^{2-\G} e^{-E/E_{cut}}\ 
\end{flalign}
with free parameters $F_0$, $\Gamma$, and $E_{cut}$.  We choose to
normalize the primary spectrum at $E_0=1$\ TeV.

%\clearpage
\begin{deluxetable}{lcccc}
\tabletypesize{\scriptsize}
\tablecaption{Parameters for cascade calculation in figures.  Parameters that 
are varied within each figure are labeled as ``v''.}
\tablewidth{0pt}
\tablehead{
\colhead{Parameter} & 
\colhead{Fig.\ \ref{sed0229_taylorlong}} &
\colhead{Fig.\ \ref{sed0229_taylor}} &
\colhead{Fig.\ \ref{sed0229_kachel}} &
\colhead{Fig.\ \ref{sed0229_arlen14}} 
}
\startdata
$\G$ & 1.2 & 1.2 & 0.67 & 1.3 \\
$E_{cut}$ [TeV] & 5.0 & 5.0 & 20 & 1\\
$F_0$ [$10^{-12}\ \erg\ \cm^{-2}\ \s^{-1}$] & $4.6$ & 4.6 & 1.1 & 12 \\
$B_{IG}$ [G] & v & v & $10^{-17}$ & 0.0 \\
$L_B$ [Mpc] & 1.0 & 1.0 & 1.0 & 1.0 \\
$t_{blazar}$ [yr] & $H_0^{-1}$ & 3 & v & $H_0^{-1}$ \\
EBL model & KD10\tablenotemark{a} & KD10\tablenotemark{a} & F08\tablenotemark{b} & F10\tablenotemark{c}
\enddata
\tablenotetext{a}{\citet{kneiske10}}
\tablenotetext{b}{\citet{franceschini08}}
\tablenotetext{c}{\citet{finke10_EBLmodel}}
\label{cascade_param_table}
\end{deluxetable}
%\clearpage

\begin{figure}
%\vspace{2.2mm} 
\vspace{1.5cm} 
\epsscale{1.0}
\plotone{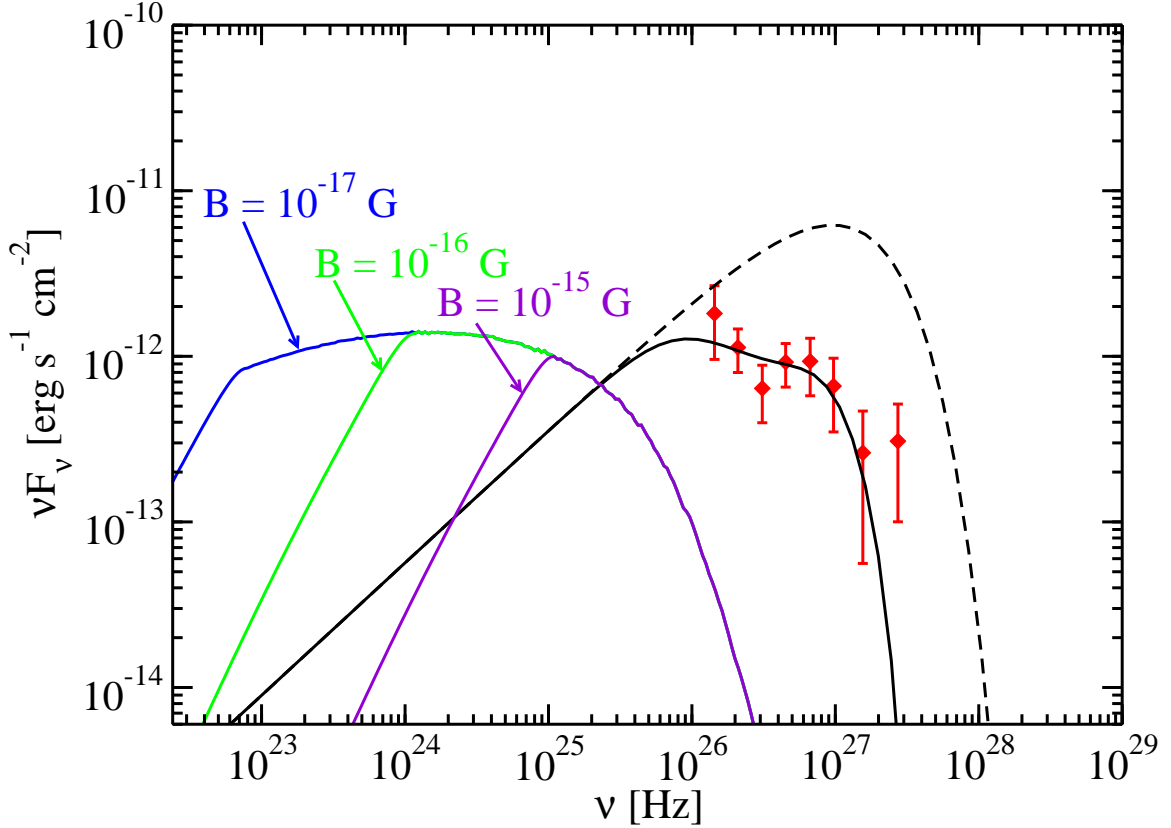}
\caption{ The cascade spectrum of 1ES~0229+200, calculated with
parameters similar to those of the calculation shown in Fig.\ 7,
first panel of \citet{taylor11}.  The HESS spectrum
\citep{aharonian07_0229} is shown as the diamonds, the primary,
unabsorbed spectrum is shown as the dashed line, and the primary
absorbed spectrum as the solid line going through the HESS data
points.  The cascade spectra are labeled by the IGMF strength used in
the calculation.  Other parameters are shown in Table
\ref{cascade_param_table}. }
\label{sed0229_taylorlong}
\vspace{2.2mm}
\end{figure}

\begin{figure}
%\vspace{2.2mm}
\vspace{1.5cm}  
\epsscale{1.0}
\plotone{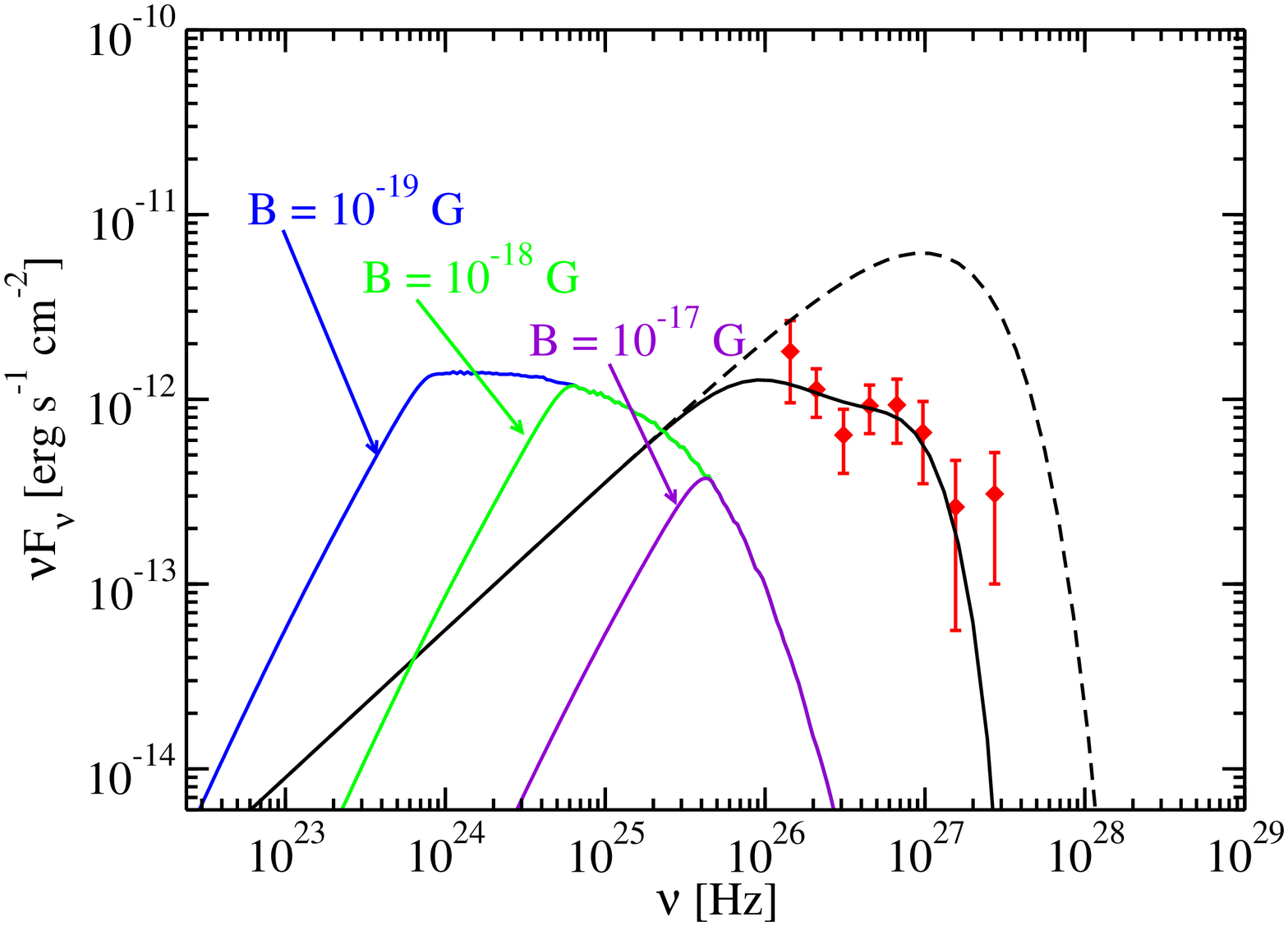}
\caption{ The cascade spectrum of 1ES~0229+200, calculated with
parameters similar to those of the calculation shown in Fig.\ 7,
second panel of \citet{taylor11}.  The HESS spectrum
\citep{aharonian07_0229} is shown as the diamonds, the primary,
unabsorbed spectrum is shown as the dashed line, and the primary
absorbed spectrum as the solid line going through the HESS data
points.  The cascade spectra are labeled by the IGMF strength used in
the calculation.  Other parameters are shown in Table
\ref{cascade_param_table}. }
\label{sed0229_taylor}
\vspace{2.2mm}
\end{figure}

In Figures \ref{sed0229_taylorlong} and \ref{sed0229_taylor} we
attempt to reproduce several of the results of \citet{taylor11} with
our simple cascade calculation.  Figure \ref{sed0229_taylorlong} can
be compared with the first panel of Figure 7 of \citet{taylor11}.
This calculation assumes that the blazar has been producing the
primary $\gamma$-ray spectrum for the age of the universe.  The
cascade results appear to be in very good agreement.  In Figure
\ref{sed0229_taylor} we perform a calculation with parameters similar
to the second panel of Figure 7 from \citet{taylor11}, where the
blazar has been producing VHE $\g$ rays for 3 years.  In this case,
the agreement is less good.  In general, our calculation predicts
lower emission at the lower energies than the MC of
\citet{taylor11}.  We conclude that our results are thus very
conservative, since they under-predict the cascade from more detailed
calculations.

\begin{figure}
%\vspace{2.2mm} 
\vspace{1.5cm} 
\epsscale{1.0}
\plotone{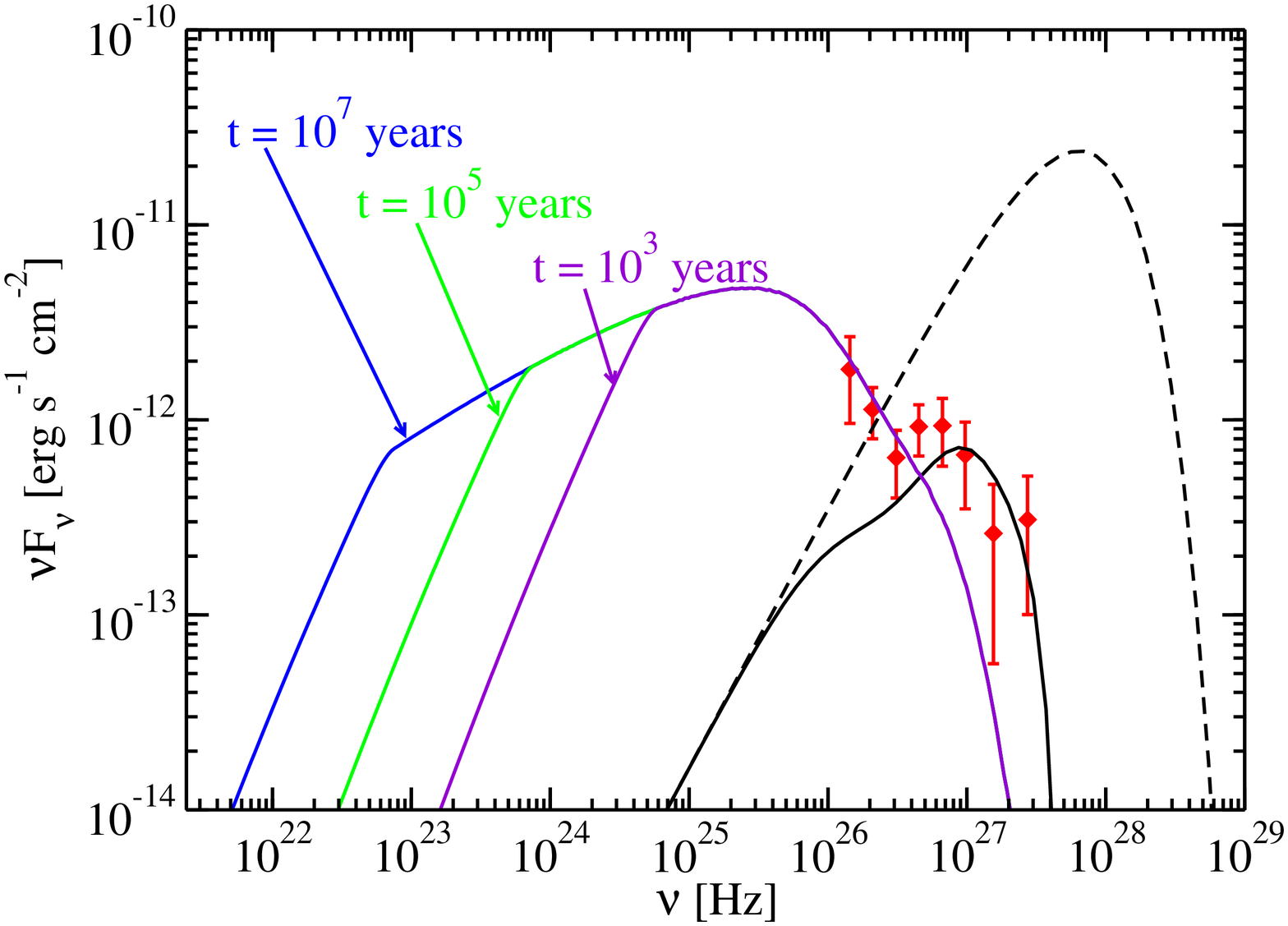}
\caption{ The cascade spectrum of 1ES~0229+200, calculated with
parameters similar to those of the calculation shown in the right
panel of Fig.\ 3 of \citet{kachelreiss12}.  The HESS spectrum
\citep{aharonian07_0229} is shown as the diamonds, the primary,
unabsorbed spectrum is shown as the dashed line, and the primary
absorbed spectrum as the solid line going through the HESS data
points.  The cascade spectra are labeled by the jet time used in the
calculation.  Other parameters are shown in Table
\ref{cascade_param_table}. }
\label{sed0229_kachel}
\vspace{2.2mm}
\end{figure}

In Figure \ref{sed0229_kachel} we produce a cascade calculation with
the same parameters as those from the right panel of Figure 3 of
\citet{kachelreiss12}.  Again, our calculation under-predicts the
lower energy emission compared to their detailed calculations, which
implies that our results are very conservative.

\begin{figure}
%\vspace{2.2mm} 
\vspace{1.5cm} 
\epsscale{1.0}
\plotone{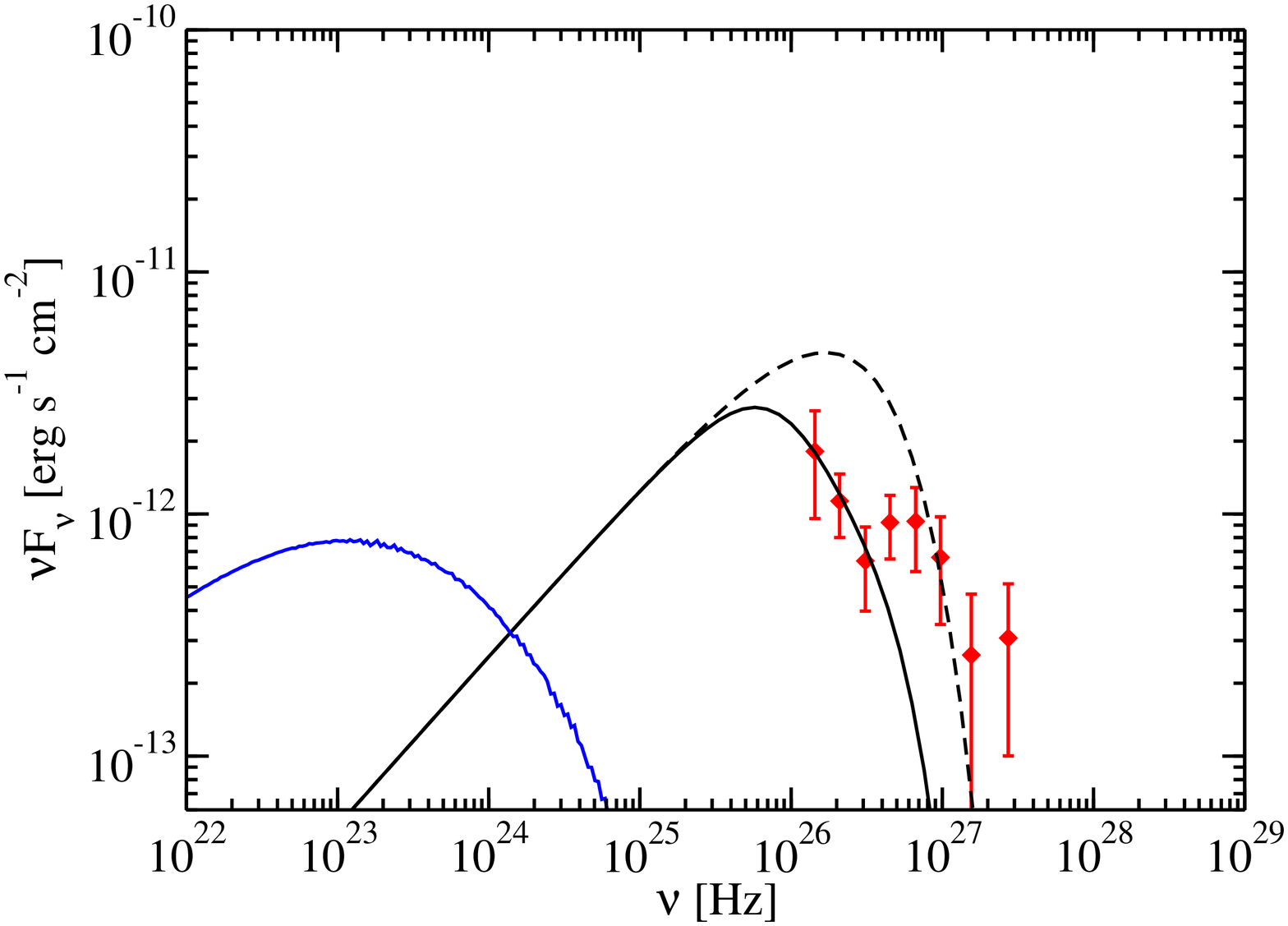}
\caption{ The cascade spectrum of 1ES~0229+200, calculated with
parameters similar to those of the calculation shown in the right
panel of Fig.\ 9 of \citet{arlen14}.  The HESS spectrum
\citep{aharonian07_0229} is shown as the diamonds, the primary,
unabsorbed spectrum is shown as the dashed line, and the primary
absorbed spectrum as the solid line going through the HESS data
points.  The cascade spectra are labeled by the jet time used in the
calculation.  Other parameters are shown in Table
\ref{cascade_param_table}. }
\label{sed0229_arlen14}
\vspace{2.2mm}
\end{figure}

In Figure \ref{sed0229_arlen14} we reproduce Figure 9 from
\citet{arlen14}, who test $B=0$.  Our cascade calculation is similar
to theirs, but a bit lower by a factor of $\approx 1.4$.

%\clearpage

%\clearpage

%\clearpage

%\clearpage

\section{Properties of PDF}
\label{pdfappendix}

In this appendix, we explore two properties of the PDF given by
Equation (\ref{pdf_gammanormal}), 
\begin{flalign}
\label{pxy}
p(x, y) & = \frac{ x^{\alpha_x-1} 
e^{-x/\beta_x} }{\G_f(\alpha_x)\beta_x^{\alpha_x} } \ 
\frac{1}{\sigma_{y} \sqrt{2\pi(1-\rho)} } \ 
\\ \nonumber & \times
\exp\left\{ -\frac{(y-\mu_y)^2}{2(1-\rho^2)\sigma_{y}^2} + 
            \frac{ \rho(x-\mu_x)(y-\mu_y)}{(1-\rho^2)\sigma_x\sigma_y} - 
	    \frac{ \rho^2(x-\mu_x)^2}{ 2 (1-\rho^2) \sigma_x^2 } \right\} \ ,
\end{flalign}
where the notation has been simplified by letting $F_{LAT}\rightarrow
x$ and $\G \rightarrow y$.  

First we show that Equation (\ref{pxy}) reduces to a bivariate normal
distribution for $\alpha_x = (\mu_x/\sigma_x)^2 \gg 1$ and $x$ is
close to $\mu_x$.  We begin by making use of the Sterling
Approximation,
\begin{flalign}
\G(\alpha) \approx \sqrt{\frac{2\pi}{\alpha}}\left(\frac{\alpha}{e}\right)^\alpha,\; \;
\alpha\gg 1
\end{flalign}
which leads to 
\begin{flalign}
\label{pxy2}
p(x,y) & = \frac{1}{2\pi\sigma_x\sigma_y\sqrt{1-\rho^2}} 
\left(\frac{x}{\mu_x}\right)^{\alpha_x-1}
\exp\left\{\frac{\mu_x^2-x\mu_x}{\sigma_x^2}\right\}\ 
\nonumber \\ & \times
\exp\left\{ -\frac{(y-\mu_y)^2}{2(1-\rho^2)\sigma_{y}^2} + 
            \frac{ \rho(x-\mu_x)(y-\mu_y)}{(1-\rho^2)\sigma_x\sigma_y} - 
	    \frac{ \rho^2(x-\mu_x)^2}{ 2 (1-\rho^2) \sigma_x^2 } \right\} \ .
\end{flalign}
The Taylor expansion of the natural logarithm is
\begin{flalign}
\ln(1 + x) = x - \frac{x^2}{2} + \frac{x^3}{3} - \ldots
\end{flalign}
which can be rewritten
\begin{flalign}
\left( 1 + \frac{x}{a}\right)^a = \exp\left\{ x - \frac{x^2}{2a} + \frac{x^3}{3a^2} - \ldots\right\}
\end{flalign}
for $x/a \ll 1$.  Using this one can show, with some algebraic
manipulation, that 
\begin{flalign}
\left(\frac{x}{\mu_x}\right)^{\alpha_x-1} 
& \approx \left(\frac{x}{\mu_x}\right)^{\alpha_x} 
= \left( 1 - 1 + \frac{x}{\mu_x}\right)^{\alpha_x}
\nonumber \\ 
& \approx \left( 1 + \frac{(x-\mu_x)\alpha_x}{\mu_x\alpha_x}\right)^{\alpha_x}
= \left( 1 + \frac{(x-\mu_x)\mu_x}{\sigma_x^2\alpha_x}\right)^{\alpha_x}
\nonumber \\ 
& \approx \exp\left\{ \frac{(x-\mu_x)\mu_x}{\sigma_x^2} 
- \frac{(x-\mu_x)^2\mu_x^2}{2\sigma_x^4\alpha_x} + \ldots \right\}
\nonumber \\ 
& \approx \exp\left\{ \frac{x\mu_x-\mu_x^2}{\sigma_x^2} - \frac{(x-\mu_x)^2}{2\sigma_x^2}
 \right\}\ .
\end{flalign}
Substitution of this into Equation (\ref{pxy2}) results in
\begin{flalign}
p(x,y) & \approx \frac{1}{2\pi\sigma_x\sigma_y\sqrt{1-\rho^2}} 
\nonumber \\  & \times
\exp\left\{ -\frac{ (x-\mu_x)^2}{ 2 (1-\rho^2) \sigma_x^2 }
             -\frac{(y-\mu_y)^2}{2(1-\rho^2)\sigma_{y}^2} + 
            \frac{ \rho(x-\mu_x)(y-\mu_y)}{(1-\rho^2)\sigma_x\sigma_y} 
	     \right\} \ 
\end{flalign}
which is the bivariate normal distribution.  

Next, we show that the PDF, Equation (\ref{pxy}) is normalized to unity.  That is, 
\begin{flalign}
\label{normalize}
1 = \int_{-\infty}^\infty \int_0^\infty\ dy\ dx\ p(x,y)\equiv J\ .
\end{flalign}
We begin by performing the integral over $y$, 
\begin{flalign}
J_y = \int_{-\infty}^\infty\ dy\ 
\exp\left\{ \frac{-(y-\mu_y)^2}{2(1-\rho^2)\sigma_y^2} + 
\frac{ \rho(x-\mu_x)(y-\mu_y)}{(1-\rho^2)\sigma_x\sigma_y} \right\}\ .
\end{flalign}
We make the substitution
\begin{flalign}
\frac{(y-\mu_y)^2}{2(1-\rho^2)\sigma_y^2} = u^2, 
\end{flalign}
giving $dy = \left[2(1-\rho^2)\right]^{1/2}\sigma_y\ du$ and
\begin{flalign}
J_y = \sigma_y\left[2(1-\rho^2)\right]^{1/2} \int_{-\infty}^\infty\ du\ 
\exp\left\{ -u^2 + 
u\ \frac{\rho(x-\mu_x)}{\sigma_x} \sqrt{\frac{2}{1-\rho^2}} \right\}\ .
\end{flalign}
This integral is tabulated, and the result is
\begin{flalign}
J_y = \sigma_y\left[2\pi(1-\rho^2)\right]^{1/2}
\exp\left\{ \frac{\rho^2(x-\mu_x)^2}{2\sigma_x^2(1-\rho^2)} \right\}\ .
\end{flalign}
Next we must do the integral over $x$.  Putting this result into 
Equation (\ref{normalize}) gives 
\begin{flalign}
J = \frac{1}{\G_f(\alpha_x)\beta_x^{\alpha_x} }
 \int_{-\infty}^\infty\ dx\ 
x^{\alpha_x-1}\  e^{-x/\beta_x} 
\end{flalign}
which is the integral of the Gamma distribution.  It is well-known to
be normalized to unity.

%!****************************************************

%\begin{thebibliography}{}

\bibliographystyle{apj}
\bibliography{3c454.3_ref,EBL_ref,references,mypapers_ref,blazar_ref,sequence_ref,SSC_ref,LAT_ref}

\end{document}